\documentclass[10pt]{article}
  \usepackage{amsmath}
 \usepackage{graphicx}
\usepackage{amssymb}
\usepackage[usenames]{color}
 \usepackage[colorlinks=true, pdfstartview=FitV, linkcolor=black, citecolor=black, urlcolor=blue]{hyperref}

\newcommand{\pause}{}

\newcommand{\D}{Deutsch}

\newenvironment{smallquotation}
               {\begin{quotation}\small}{\normalsize\end{quotation}}

     \newcommand{\mc}[1]{\ensuremath{\mathcal{#1}}}

\newcommand{\be}{\begin{displaymath}}
\newcommand{\ee}{\end{displaymath}}

\usepackage{latexsym}

\begin{document}
\title{Decisions, Decisions, Decisions: Can Savage  Salvage Everettian Probability?\thanks{A slightly shorter version of this paper is to appear in a volume edited by Jonathan Barrett, Adrian Kent, David Wallace and Simon Saunders, containing papers presented at the \emph{Everett@50} conference in Oxford in July, 2007, and the \emph{Many Worlds@50} meeting at the Perimeter Institute in September, 2007. See Price (2007) for the original talk on which this paper is based.}}
\author{Huw Price\thanks{Email: \href{mailto:huw@mail.usyd.edu.au}{huw@mail.usyd.edu.au}}}
\date{\today}

\maketitle
 \begin{abstract}
\noindent Critics object that the Everett view cannot make sense 
of quantum probabilities, in one or both of two ways:  either it cannot make sense of probability at all, or it cannot explain why probability should be governed by the Born rule. David Deutsch has attempted to meet these objections by appealing to an Everettian version of Savage's rational decision theory.  Deutsch  argues not only that an analogue of classical decision under uncertainty makes sense in an  Everett world; but also that under reasonable assumptions, the betting odds of a rational Everettian agent should be constrained by the Born rule. Deutsch's proposal has been defended and developed by David Wallace, and in a different form by Hilary Greaves. In this paper I offer some objections to the Deutsch-Wallace-Greaves argument, focussing in particular on the supposed analogy with classical decision under uncertainty.  \end{abstract}

\section{Fabulous at fifty?}

Our forties are often a fortunate decade. Understanding ourselves better, resolving old problems,  we celebrate our fiftieth year with new confidence, a new sense of purpose. So, too,  for the Everett interpretation, at least according to one reading of its recent history. Its fifth decade has resolved a difficulty that has plagued it since youth, indeed since infancy, the so-called problem of probability. 
Critics have long maintained that the Everett view cannot provide an adequate account of quantum probabilities, in one or both of two ways: either it cannot make sense of probability at all, in a world in which `all possibilities are actualised'; or, at best, it cannot explain why probability should be governed by the Born rule. 

According to the optimistic view just mentioned, these problems have been sorted out, in the past decade, by an approach due to David Deutsch (1999). \D\  argues not only that an analogue of decision under uncertainty (of the kind traditionally associated with probability) makes sense in an  Everett world; but also that under reasonable assumptions, the betting odds of a rational Everettian agent should be constrained by the Born rule. In one important respect, \D\ argues, 
 probability is actually in \emph{better} shape in the Everettian context than in the classical `one-branch' case. If so, the Everett view turns fifty in fine form indeed.

\D 's argument has been reformulated and clarified by David Wallace (2002, 2003, 2006, 2007). Wallace stresses  the argument's reliance on the distinguishing symmetry of the Everett view, viz., that all possible outcomes of a quantum measurement  are treated as equally real. The argument thus makes a virtue of what is usually seen as the main obstacle to making sense of probability in this context. Further important contributions have been made by several other writers, such as Simon Saunders (1998, 2005) and Hilary Greaves (2004, 2007a, 2007b). The result is a fascinating collection of papers, of great interest to researchers in philosophy of probability, as well as the foundations of QM.

But does it work? In this paper, presenting the pessimistic face of turning fifty,   I want to argue that it does not. I'll argue in particular that the distinguishing symmetry of the Everett view (the fact that it treats all outcomes of a quantum measurement  as equally real) is the downfall rather than the saviour of   the \D -Wallace (DW) argument (or  the \emph{Oxford} approach, as I'll sometimes call it). For Everett's  new ontology  provides something new for agents to have preferences \emph{about;}  and this, as we'll see, makes it inevitable that rational Everettian agents need not be \D ian agents. (In other words, rationality does not require that their decision behaviour be modelled by an analogue of classical decision under uncertainty, with the Born weights playing the role of subjective probability.) And I'll argue that this is no mere abstract possibility. Once the loophole is in view, it is easy to see why we might prefer not to be \D ian agents, at least in some respects, if we believed we lived in an Everett world.

Little  in this paper is new. As I'll note, the main point is made by Greaves herself (though I think she underestimates its significance). But the package has not previously been assembled in this form, so far as I know. 

\section{A missing link?}

It is often suggested that the problem of probability in the Everett interpretation is analogous to, and perhaps no worse than, a problem long-recognised  in the classical case. An early and forceful version of this claim is that of David Papineau (1996). Papineau considers what he terms the Decision-Theoretical Link:

\begin{quote}
\emph{The Decision-Theoretical Link.} We base rational choices on our knowledge of objective 
probabilities. In any chancy situation, a rational agent will consider the 
difference that alternative actions would make to the objective probabilities of desired results, and then opt for that action which maximizes 
objective expected utility. (1996, 238)
\end{quote}

\noindent `Perhaps surprisingly,' Papineau continues, `conventional thought provides no agreed further justification [for this principle]':

\begin{quote}
Note in this connection that what agents want from their choices are desired \emph{results,} rather than results which are objectively \emph{probable} (a choice that makes the results objectively probable, but 
unluckily doesn't produce them, doesn't give you what you \emph{want}). This 
means that there is room to ask: \emph{why} are rational agents well advised to 
choose actions that make their desired results objectively probable? However, there is no good answer to this question \dots . Indeed many philosophers in this area now simply take it to be a 
primitive fact that you ought to weight future possibilities according to 
known objective probabilities in making rational decisions.  \ldots\  It 
is not just that philosophers can't agree on the right justification; many 
have concluded that there simply isn't one. (1996, 238)
\end{quote}

Applying this to the many worlds case, Papineau suggests that the Everettian is therefore entitled simply to \emph{assume}  that rational decision in a branching universe is constrained by certain physical magnitudes postulated by QM: 

\begin{quote}
[T]he many minds view should simply stipulate that the quantum 
mechanical coefficients \ldots\   provide a decision-theoretic basis for rational decisions. As to a
justification for these stipulations, the many minds theory can simply 
retort that it provides as good a justification as conventional thought 
does for treating its probabilities similarly---namely, no good justification 
at all. (1996, 238--239)
\end{quote}

Supporters of the DW approach claim that if their argument succeeds, we actually get more than this: we get a \emph{proof} that rational decision in an Everett world is properly constrained by the Born rule. But if the proof doesn't quite go through, then no matter -- we are still no worse off than in the classical case. 

In my view,  this comparison both exaggerates the difficulty in the classical case, and misrepresents and hence underestimates the difficulty in the Everettian case. I want to argue that so long as we keep in mind -- as emphasised by the \emph{pragmatist} tradition in philosophy of probability\footnote{A.k.a.~the \emph{subjectivist} tradition, though I think there are good reasons to prefer the former term.} --  that probability properly \emph{begins} with 
decision under uncertainty, there isn't any pressing mystery about 
Papineau's Decision-Theoretic Link. (There may be a problem for some views of probability, but those views are far from compulsory -- and the problem itself counts against them, in so far as it is a problem.) On the other side, the crucial issue for the Everettians is in a sense \emph{prior} to probability:  it is 
the problem whether decision under uncertainty -- or some suitable analogue -- 
makes any sense in the many worlds case. This problem is prior, in particular, to the question as to whether we might take it to be a brute fact that rational Everettian decision is constrained by the Born rule; for until it is clear that there is something coherent to constrain in this way, such an assumption is premature, if not question-begging. The true problem in this vicinity for the Everett view -- somewhat misleadingly labelled as the problem of probability, and with no analogue in the classical case -- is the problem about uncertainty.

To give a sense of the pragmatist viewpoint, imagine creatures who make maps of their surroundings,  marking the positions of various features of practical significance: blue dots and lines where they find the liquids they drink and wash in, for example, and green shopping baskets where they find things to eat. 

Is there any mystery, or primitive assumption at work, in the fact that the mapmakers use these maps as a guide to where they 
can drink, wash and eat? No, for that's precisely what the maps are \emph{for.}  (To understand the map is to understand that this is how it is \emph{used,} as a Wittgensteinian might say.) The mapmakers might find other interesting questions in the vicinity, of course. Why are blue 
lines correlated with contour lines in such a distinctive way? Is  
there is any unified \emph{physical} account, either  of the stuff they drink and wash 
in, or of the stuff they eat? And perhaps most interestingly, what is it about the mapmakers and their environment that explains why this mapmaking practice is both \emph{possible} and \emph{useful,} to the extent that it is? But these are not the practical puzzle about why someone who adopts the map should use it to guide her drink-seeking and food-seeking behaviour: to adopt the map \emph{is} to take it as a practical guide. 

A pragmatist regards probabilistic models in this same practical spirit. They are maps  to guide us in making decisions under uncertainty, in particular domains. The previous map said, `Go \emph{here} if you need a wash or a drink, \emph{there} for something to eat.' This map says, `Use \emph{this number} if you need a credence -- i.e.,   if you have to make a decision with imperfect knowledge about something that matters.' If  a probability map just \emph{is} a practical guide to decision under uncertainty,  it isn't a mystery why it should be  used for 
exactly that purpose. There's no primitive assumption, and no decision-theoretic `Missing Link'. Again, there might be other interesting questions in the vicinity -- e.g., about whether, 
or why, decision-theoretic `blue lines' are correlated with something 
else on our maps -- but these are not the practical puzzle about why 
probability properly guides action.\footnote{Would it matter is there wasn't any unified story about `something 
else' that correlated with decision theoretic probability -- if `probability' turned out to be more like `food' (where there isn't any interesting natural kind, apparently), than like `liquid for washing and drinking' (where there is a natural kind, at least to a first approximation)? If so, why? Obviously there is  (much) more to be said on this matter, but I hope the food example goes some way to undermining the rather simplistic conception of the options that tends to characterise objections to the pragmatist viewpoint. Wallace, for example, says, `Whilst it is coherent 
to advocate the abandonment of objective probabilities, it seems implausible: 
it commits one to believing, for instance, that the predicted decay rate of radioisotopes is \emph{purely} a matter of belief.' (2002, 15, my emphasis) Would abandonment of the view that `food' picks out a natural kind commit us to believing that it is \emph{purely} a matter of taste? Surely there's room for some subtlety here -- room for a bit of natural science, in effect, that begins by asking why the concepts concerned (`food' or `probability') should be useful for creatures in our circumstances, to the extent that they are useful (and anticipates an answer that refers both to aspects of the environment and to aspects of ourselves)?}

The pragmatist thus applauds the Oxford approach for beginning with the question of rational decision,  but asks it to allow that classical probability does the same. With that concession in place, the issues line up as they should. There's no essential primitive assumption in the classical case, to which Everettians can appeal in the spirit of `we do no worse than that'. And the crucial question, the crux of the so-called problem of probability, is whether the Everett picture allows any analogue of decision-theoretic uncertainty -- any analogue of \emph{credence,} in effect.\footnote{To the satisfaction of what \emph{need} could Everettian probability be our guide, as it were?}

\section{Rational choice from Savage to \D}

In a recent survey article, Hilary Greaves provides a concise summary of the classical model of rational decision theory due to Savage, and of its modification by \D\ in the Everettian case. Here I follow and appropriate Greaves' exposition, compressing further where possible:

\begin{quote}
Decision theory is a theory designed for the analysis of 
rational decision-making under conditions of uncertainty. One considers an agent who is 
uncertain what the \emph{state} of the world is: for example, she is uncertain whether or not it 
will rain later today. The agent faces a choice of \emph{acts:} for example, she is going for a 
walk, and she has to decide whether or not to take an umbrella. She knows, for each 
possible state of the world and each possible act, what will be the \emph{consequence} if the state 
in question obtains and the act in question is performed: for example, she knows that, if 
she elects not to take the umbrella and it rains, she will get wet. The agent is therefore 
able to describe each of her candidate acts as a function from the set of possible States to 
the set of Consequences. (Greaves 2007b, 113)
\end{quote}

As Greaves explains, Savage then shows that given certain plausible rationality constraints, one may prove a `representation theorem':

\begin{quote}
[F]or any 
agent whose preferences over acts satisfy the given rationality constraints, there exists a 
unique probability measure $p$ on the set of States, and a utility function $U$ on the set of 
Consequences (unique up to positive linear transformation), such that, for any two acts $A, 
B,$ the agent prefers $A$ to $B$ iff the expected utility of $A$ is greater than that of $B$. Here, the 
expected utility of an act $A$ is defined by: 
\be
EU(A) \mathrel{\mathop:}= \sum_{s \in \mc{S}} p(s)\cdot U(A(s)).\ee
\end{quote}

\D\ modifies this classical apparatus by replacing the set of States with the set of branches that will result from a specified quantum measurement, and the set of Consequences with, as Greaves puts it `things that happen to individual copies of the agent, \emph{on particular branches.}' An Act is still a function from States to Consequences, but in this new sense of Consequence: in other words, it is an assignment of Consequences or rewards to branches defined by measurement outcomes. Then, by  `imposing a set of rationality constraints on agents' preferences among such quantum 
games',

\begin{quote}
Deutsch is able to prove a representation theorem that is analogous in many 
respects to Savage's: the preferences of a rational agent are representable by a probability 
measure over the set of States (branches) for every possible chance setup, and a utility 
function on the set of Consequences (rewards-on-branches), such that, for any two 
Everettian acts $A, 
B,$ the agent prefers $A$ to $B$ iff $EU(A)$$>$$EU(B).$ (Expected utility is 
defined via the same formal expression as above.) (2007b, 115)
\end{quote}

\subsection{Wallace's version of \D 's approach}

How does an analogue of uncertainty enter the picture? Since much hangs on this issue, I'll reproduce a  clear account of the relevant move from Wallace (2007, 316--317), in a formulation of the axioms in which uncertainty is `on the surface'.
\begin{quotation}
\noindent [D]efine a 
\emph{likelihood ordering}
as some two-place relation holding between ordered pairs $\langle E,M\rangle$, where 
$M$ is a quantum measurement and $E$ is an event in $\mc{E}_M$ (that is, $E$ is a subset of the
possible outcomes of the measurement). We write the relation 
as $\succeq$:
\be E|M \succeq F|N \ee
is then to be read as ``It's at least as likely that some outcome in $E$ will obtain 
(given that measurement $M$
is carried out) as it is that some outcome in $F$ will obtain (given that measurement $N$
is carried out)''. We define $\simeq$ %and $\succ$ 
\dots\ as follows: $E|M\simeq F|N$ if
$E|M\succeq F|N$ and $F|N\succeq E|M$
\ldots .

We will say that such an ordering is \emph{represented} by a function $\Pr$ from pairs $\langle E,M\rangle$
to the reals 
if
\begin{enumerate}
\item $\Pr(\emptyset|M)=0$, and $\Pr(\mc{S}_M|M)=1$, for each $M$.
\item If $E$ and $F$ are disjoint then $\Pr(E\cup F|M)=\Pr(E|M)+\Pr(F|M)$.
\item $\Pr(E|M)\geq\Pr(F|N)$ iff $E|M \succeq F|N$.
\end{enumerate}
The ordering is \emph{uniquely represented} iff there is only one such $\Pr$.

The subjectivist program then seeks to find axioms for $\succeq$ so that any agent's preferences
are uniquely represented. \ldots\ [I]n a \emph{quantum-mechanical} context we can manage with a set of axioms which
is both extremely weak  \ldots\ and fairly simple.  To state them,
it  will be
convenient to  define a \emph{null event}: an event $E$ is 
\emph{null with respect to M} (or, equivalently, $E|M$ is null) iff $E|M \simeq \emptyset|M$. 
(That is: $E$
is certain not to happen, given $M$).
If it is clear which $M$ we're referring to, we will sometimes drop the $M$ and refer to $E$ as 
null \emph{simpliciter}.

We can then say
that a likelihood ordering is \emph{minimally rational} if it satisfies the
following axioms:
\begin{description}
\item[Transitivity] $\succeq$ is transitive: if $E|M\succeq F|N$ and $F|N \succeq G|O$, 
then $E|M\succeq G|O$.
\item[Separation]
There exists some $E$ and $M$ such that $E|M$ is not null.
\item[Dominance]
If $E \subseteq F$, then $F|M \succeq E|M$ for any $M$, with $F|M \simeq
E|M$ iff $E-F$ is null.
\end{description}

This is an extremely weak set of axioms for qualitative likelihood  \ldots . Each, translated into words, should be immediately
intuitive:
\begin{enumerate}
\item Transitivity: `If A is at least as likely than B and B is at least as likely
than C, then A is at least as likely than C.'
\item Separation: `There is some outcome that is not impossible.'
\item Dominance: `An event doesn't get less likely just because more
outcomes are added to it; it gets more likely iff the outcomes which are added are not themselves
certain not to happen.' 
\end{enumerate}
\end{quotation}

\subsection{Comments}
Wallace is right, of course, that any reasonable notion of qualitative  likelihood should be expected to satisfy these axioms. But to characterise the target is not yet to hit the bullseye. The axiom system provides no answer to the puzzle as to what such a notion of subjective likelihood could amount to, in the Everettian context, where there is no classical uncertainty. So if we assume these axioms as the basis for a quantum version of the representation theorem, we are \emph{assuming,} rather than \emph{demonstrating,} that there is some non-trivial Everettian analogue of decision-theoretic uncertainty. (We know that there is a \emph{trivial} measure of likelihood available in the Everett world: it assigns equal and maximal likelihood to all results not certain not to happen. But this measure doesn't satisfy Dominance,  for it fails the requirement that Wallace glosses as: `An event \ldots\ gets more likely iff the outcomes which are added are not themselves certain not to happen.')

I'm not suggesting that Wallace is confused about these points, of course. On the contrary, he and others have discussed at length what the required analogue of uncertainty might be. (More on this below.) Nevertheless, I think it is worth stressing that the availability of an appropriate notion of uncertainty doesn't \emph{emerge from} the DW argument, but is \emph{presupposed by it.} So if we sceptics are challenged to say which of these axioms we disagree with, we have at least the following answer: we're sceptical about any axiom -- e.g., Wallace's Dominance, for one -- that presupposes an analogue of uncertainty. And we'll remain sceptical, until our opponents convince us that they have a notion that will do the job.

\section{The quest for uncertainty}

There are two main proposals as to how we might make sense of the required notion of decision-theoretic uncertainty, or likelihood. The first turns on the claim that an Everettian agent  properly feels genuine though \emph{subjective} uncertainty, as to which branch she will find herself in, after a quantum measurement. The second argues that the required analogue of uncertainty isn't really any sort of uncertainty at all, but rather what Greaves calls a `caring measure' -- a measure of how much `weight' a rational agent should give to a particular branch.

Greaves (2004) herself has offered some robust and (to me) fairly persuasive criticisms of the subjective uncertainty approach. I want to raise an additional difficulty for this view, not mentioned (so far as I know) in the existing literature. It, too, seems to suggest that Greaves' approach is the right approach, if anything is -- though it also contains the seeds of an objection to her  view, I think. It turns on what I hope are some uncontroversial observations about the role of subjects and survival in rational decision theory.

\subsection{Decision theory doesn't care about personal identity}

First, rational decision theory depends on the fact that agents care about outcomes, but not on whether these outcomes are things that the agent herself experiences. They are things the agent has preferences \emph{about} -- her \emph{loci of concern,} as I'll say --  but her own future experiences are just one possible class of such things. 

Thus consider a case in which the relevant outcome is that a child is happy at a specified time in the future. What is revealed by my betting preferences is the utility -- for \emph{me, now} -- of the child's being happy, not the utility for the child (now or in the future). Similarly if we replace the child by some future person-stage of me: what's revealed by my betting preferences \emph{now} is the utility \emph{for the present me} of the future me being happy, not the utility \emph{for} the future me.

In general, then, it is entirely inessential to classical subjective decision theory that its subjects \emph{have} future selves, at the time of the relevant Consequences. We can imagine a race of short-lived creatures for whom all decisions would be like death-bed decisions are for us --  we prefer that there should be some futures rather than others, and act so as to maximise our expected utility, in the light of uncertainty,  of a future that we ourselves shall not experience. Indeed, we can imagine a view of the metaphysics of personal identity according to which this is inevitably our own situation: there is no genuine trans-temporal personal identity, according to this view, and our decisions cannot but concern a future that we ourselves will never see. In such a case, of course, there is simply no place for subjective uncertainty about which future we will experience. Properly informed by our metaphysical friends, we \emph{know} that we will experience none of them. No matter -- our classical decision theory takes all this in its stride. (The relevant uncertainty is just the agent's uncertainty as to which future is \emph{actual.})

I've emphasised these points to highlight a respect in which the subjective uncertainty  version of  \D ian decision theory seems inevitably less general than its classical ancestor. It needs to take sides on some heavy-duty metaphysical issues about personal identity, so as to rule out the view according to which there is no such thing as genuine survival. And even with the help of such metaphysics, it only applies, at least directly, to cases in which the agent does survive -- for it is only in these cases that the relevant notion of subjective expectation makes sense. On my death bed, I expect nothing for tomorrow --  Everett-embarrassing Lewisean loopholes to one side, at least!\footnote{See Lewis 2004. The claim I make here about the generality of the subjective uncertainty view might be challenged. Wallace himself says:

 \begin{quote}
 `Subjective' should not be taken too literally here. The subjectivity lies 
in the essential role of a particular location in the quantum universe (uncertainty isn't visible from a God's-eye view). But it need not be linked to first-person expectations: `there will be a sea battle tomorrow' might be 
as uncertain as `I will see spin up'. (2007, 314)
\end{quote}
\noindent I don't have space to pursue this point here, but it seems that for a view not linked to first-person expectations, the subjective uncertainty approach spends an inordinate amount of time discussing issues of personal identity over time. My point is that these issues are essentially irrelevant to classical subjective decision theory, for which the only `I' who matters is the `I' at the time of decision.}

\subsection{What does survival mean, and why does it matter?}

There other difficulties lurking nearby. For what does \emph{survival} mean? Consider a classical analogy. Suppose, modifying the above example, that I am making a decision which will affect the happiness of a group of children. There is no uncertainty involved. I have several options, each with different consequences for the happiness of each of the children. I act so as to maximise \emph{my} utility, which  (by assumption) depends on nothing other than (my beliefs about) the resulting happiness of the children. There are many ways my utility function might vary over the $N$-dimensional space representing the degree of happiness of each of the $N$ children, but suppose I am a simple-minded consequentialist -- then my utility just goes by the total happiness, summed over the group of children.

Now suppose that initially I plan to make my choice in the belief that I will not live to see the outcome, but then  learn that I am going to be reincarnated as one of the children. Does survival make any difference to my calculation? In particular, does it provide a role for an ascription of a \emph{probability} to the matter of which child I shall be?

There are two ways  it can fail to do so. One possibility is that I believe that I will find myself in a child's body with the same `global' perspective as before -- the same concern for the welfare of the group of children as a whole. If so, then although survival means that I will live to enjoy the utility of my choice, 
 it makes no difference to the calculation. My reincarnated self will have the same utility function as I do now, whichever child he turns out to be. Another possibility is that `I' will find `myself' with the preferences of the child in question, in which case `I' will be happy or sad according as he is happy or sad -- the fate of the other children won't come into it. But unless I take this \emph{now} as a reason to abandon my even-handed approach to the future welfare of the children -- more on this possibility in a moment -- then again, survival makes no difference. (After all, I knew already that there would be a child with that perspective.) In either case, then, there's no role for a probability in my calculation, despite the survival. 
In other words, if reincarnation doesn't break the symmetry of my original attitude to the children then it is irrelevant; if it does break the symmetry, but only from the moment  of reincarnation, then it is not survival, in any sense that matters here. 

In the classical case  there's a third possibility: the news about reincarnation might break the symmetry at the beginning, in the sense that I will make my initial utility calculation in a different way altogether: I'll base it on the future happiness of one child, weighted by my credence that that child is the future me. 

Notice that although this last option makes sense in the classical case -- we can intelligibly suppose that the symmetry is broken in this way -- it would be absurd to suggest that rationality \emph{requires} me to do it this way. Rationality requires that I act now on the basis of what matters to me now. This might depend on what happens to me in the future, in so far as we can make sense of that notion, but it need not do so: rationality doesn't impose that outlook. And it doesn't take \emph{my} survival to break the symmetry, either. We can get the same result by adding the supposition that only one of the group of children -- I don't know which one -- will survive to the relevant time in the future. In this case, too, I may well need a decision theory that can cope with uncertainty. But that need stems from the fact that the world breaks a symmetry among a class of things that I care about, not from the fact (if it is a fact) that one of those things is a future version of `me'.

Now transpose this to an Everett case, in which I am facing a choice between gambles on a quantum measurement. I believe that there are multiple real future branches, one (or at least one) for each possible measurement outcome. The different gambles produce different outcomes for individuals whose welfare I care about, in each of the branches. Does it matter whether any of those individuals are `me'? As in the previous case, there are two sense in which it seems not to matter: the first in which the future individual concerned has the same `global' concerns as I have, and simply takes himself to be enjoying the net global utility which I could already foresee with certainty; and the second in which the future individual is no longer the me that matters, for present purposes, in that his concerns have become more selfish than mine (he only cares about what happens in his branch).

Is there a third sense, a symmetry-breaking sense, such that my initial decision should now become a weighted sum over a range of possibilities? The subjective uncertainty view claims that there is -- that we can make sense of the symmetry being broken `subjectively', though not objectively. (The whole point of the Everett view is that it is not broken objectively.) My point is that even if this were so -- and even if it could conceivably be relevant to the decision maker, who clearly doesn't occupy any of these subjective stances, to the exclusion of any of the others -- it couldn't be a \emph{rationality} requirement that decisions be made on that basis. If an Everettian agent's present utilities depend on his view of the welfare of occupants of multiple future branches, then -- just as in the case of the children -- it is irrelevant which of those branches he takes himself (subjectively!) to occupy.

One lesson of the move to the case in which only one child would survive was that in that case, uncertainty was forced on us by the ontology -- uncertainty about which of our various loci of concern (the children) would be `actual' (alive), at the relevant time in the future.\footnote{The other lesson was that it isn't relevant whether the child in question is `me', unless my present preferences make it so.} The shift to the Everett world takes us in the other direction. It removes the uncertainty, by rendering equally real all of an agent's previous (`classical') loci of concern.  As in the case of the children, it couldn't be a requirement of rationality that we put the uncertainty back in, even if we did have a notion of subjective uncertainty that could do the job.

\section{Two cheers for the caring measure}

These considerations were intended to supplement the arguments that Greaves, in particular, has offered against the subjective uncertainty approach. It is a great advantage of Greaves' caring measure approach that it does take seriously the new `global' perspective of an Everettian agent. 
Moreover, although Greaves herself formulates the view in terms of an agent's care \emph{for her own future descendants,} this seems inessential. If the approach works, then caring measures are a rational discounting factor for any source of utility from future branches, whether it concerns the welfare of the agent's own descendants or not. (This inoculates the approach from issues about personal identity, survival, and the like -- those matters are simply irrelevant, in my view, for the reasons above.)

So why only two cheers for the caring measure? Because (I claim) its victory over the subjective uncertainty view turns on a point that proves its own Achilles' Heel. Once the new \emph{global} viewpoint of the Everettian agent is on the table, it turns out to count against Greaves' proposal, too, in two senses. First, and more theoretically, it shows us how to make sense of an Everettian agent who, although entirely rational in terms of his own utilities, fails to conform to the \D ian model.\footnote{As I noted, Greaves (2004, 451--452) is well aware of this possibility.} Second, and more practically, it leads us in the direction of reasons why we ourselves might reasonably choose  to be such agents, if we became convinced that we lived in an Everett world. 
Thus I want to make two kinds of points:
\begin{enumerate}
\item The new ontology of the Everett view introduces a new  locus of possible concern. The utilities of  a rational Everettian agent might relate directly to the global ontology of the Everett view, and only indirectly, if at all, to `in-branch' circumstances. (For such an agent uncertainty enters the picture, if at all, only in a classical  manner.) The DW argument rules out such agents by fiat, in effect, by assuming that utilities are an `in-branch' matter.

\item Even if we restrict ourselves to agents whose global preferences do show some `reasonable' regard for the welfare of their in-branch descendants, it doesn't follow that the global preference should be to maximise a Born-weighted sum of in-branch utilities. This isn't the only option, and there are at least two kinds of reasons for thinking is should not be the preferred option: one objects to \emph{summing,} and the other to \emph{weighting.} 

\end{enumerate}

\subsection{The threat of globalisation}

One of the lessons of the previous section was that `Where goes ontology, there a possible preference'. Decision theory places no constraints on what agents care about, other than that it be \emph{real.}\footnote{It would be interesting to explore the subtleties of this restriction in the case of modal realities, variously construed, but that would take us a long way astray.} The new ontology of the Everett view -- the global wave function itself -- thus brings in its wake the possibility of an agent who cares about \emph{that.} Hence the challenge, in its most general form: by what right do we assume that the preferences of Everettian agents are driven by  `in branch' preferences \emph{at all?}

This may seem a trivial point. After all, even in the classical case we can imagine agents whose preferences are so non-specific that uncertainty disappears, from their point of view. (All I care about is that something happens tomorrow.) This doesn't seem to lessen the importance of the theory for less uninteresting agents. Isn't a similar move possible in the Everettian case? Granted, the caring measure approach doesn't apply non-trivially to any possible agent, but if it applies to a large and interesting class of agents, isn't that good enough? Can't we just restrict ourselves by fiat to the case of agents whose preferences do depend on what happens in future branches?

Unfortunately for the DW argument, no such restriction seems likely to get it off the hook at this point, unless it is so strong as to be question-begging. For the argument's own prescription for rational choice between quantum games -- `Choose the option that maximises expected utility, calculated using the Born weights' -- is \emph{itself} a rule for choosing between future global wave functions. To adopt such a rule, then, is to accept a principle for choice at this global level -- at which, obviously, there is neither uncertainty nor any analogue of uncertainty. (The caring measure is already rolled up inside this global rule.) 

But this means that if someone already has a \emph{different} preference at the global level -- in particular, a \emph{different} way of ranking wave functions according to what goes in in the branches they entail (we'll meet some examples in a moment) -- then the DW argument has nothing to say to them. Rationality may dictate choice in the light of preference, but it doesn't dictate preference itself. (Recall our principle: `Where goes ontology, there goes possible preference.') \emph{Pace} Greaves (2004, 452), this argument has no analogue in the classical case, because in that case there's no ontology at the `global' level -- only the `local' ontology, and then uncertainty about that.\footnote{Could someone sufficiently realist about probability find a classical analogue in the possibility of an agent who cared \emph{directly about the probabilities,} so that probability itself provided their (classical) analogue of additional ontology provided by the Everett view? As a pragmatist, I'm inclined to say that there is a potential analogy here, and so much the worse for some kinds of realism about probability. For present purposes, however, let me just emphasise two points of disanalogy. First, the extra ontology of the Everett view is in no sense optional -- on the contrary, it is the heart of the physical theory itself. Second, what this extra ontology consists in, \emph{inter alia,}  is a whole lot more of the kind of things that ordinary people care about anyway (as opposed, so to speak, to something that only a metaphysician could love). The suggested analogy, by contrast, would fly in the face of the fact noted by Papineau, in the passage we quoted above: `[W]hat agents want from their choices are desired \emph{results,} rather than results which are objectively \emph{probable} (a choice that makes the results objectively probable, but 
unluckily doesn't produce them, doesn't give you what you \emph{want}).'} By actualising the epistemic \emph{possibilia,} the Everett view introduces a new locus of possible preference, and hence leaves itself vulnerable to this challenge.

It seems to me that at a minimum, this argument establishes that the DW argument cannot rest on a principle of \emph{rationality.} It shows us how to imagine agents whose preferences are such it would clearly be \emph{irrational} for them to follow the DW prescription, in an Everett world -- given their preferences, rationality requires them to make different choices about the global wave function.\footnote{I note in passing another reason for thinking that the DW argument might be vulnerable in this way, viz., that the argument itself appeals to attitudes to global states at a crucial point. The principle Wallace calls Equivalence requires \emph{as a matter of rationality}  that an agent should rank two Acts equally, if they give rise to the same global state. In relying on this principle, the argument can hardly afford to be dismissive about the idea of preferences for global states.}

It seems to me that the best response to this objection would be to fight back at the level of global preferences --  to try to show that a preference for anything other than maximising a Born-weighted sum of in-branch payoffs would be (in some sense) \emph{unreasonable.}   My next goal is to show that this fight will be an uphill battle: in some respects, such anti-\D ian global preferences look very reasonable indeed. Reasonable folk like us might well conclude that there are better ways to care for the welfare of future branches.

\subsection{Choices about group welfare}

The considerations I have in mind of are two kinds: 

\begin{enumerate}
\item An MEU  model seems in some respect just the \emph{wrong kind of model} for this kind of decision problem, which concerns the welfare of a group of individuals (i.e., the inhabitants of multiple future branches). In the classical case, there are well-known difficulties for weighted sum approaches to the welfare of groups of individuals, and I want to argue that similar considerations seem to apply in the Everett case.
\item In so far as an MEU model is appropriate, there is no adequate justification for discounting the interests of low-weight individuals. I'll make this point by asking why we feel entitled to discount low-weight alternatives in the classical case, and arguing that this justification has no analogue in an Everettian framework (whether via a caring measure or subjective uncertainty).
\end{enumerate}
Concerning both points, it will be helpful to have a vivid example in mind, to guide (or pump) our intuitions.

\section{Legless at Bondi}

Suppose I'm swimming at Bondi Beach, and a shark bites off my right leg. Saved from the immediate threat of exsanguination, I'm offered a wonderful new treatment. In the hospital's new Cloning Clinic (CC), surgeons will make a (reverse) copy of my left leg, and attach it in place of the missing limb. As with any  operation there are risks: I might lose my left leg, or die under anaesthetic. But if these risks are small, it seems rational to consent to the procedure.

On my way to the CC, however, I learn a disturbing further fact: it is actually a \emph{Body} Cloning Clinic (BCC). The surgeons are going to reverse-copy all of me, remove the good right leg from the clone, and attach it to me. So I get two good legs at the expense of a legless twin. (This poor chap wakes up, complaining,  in another ward -- and is handed the consent form, on which `he' accepted the risk that he would find himself in this position.) 

This new information seems to make a difference. I'm a lot less happy at the thought of gaining a new leg at this cost to someone else -- especially someone so dear to me! -- than I was at gaining it merely at the cost of risk to `myself'. What's more, my discomfort seems completely insensitive to considerations of subjective uncertainty or branch weight. It isn't lessened by the certainty -- as I told the story -- that it would be `me' who gets the legs; nor, apparently, by the information that the other ward is in another branch of the wave function, with very low weight. 

Perhaps these intuitions simply need to be stared down, or massaged away. Perhaps I would lose them, once fully acquainted with the meaning of probability and uncertainty, or their analogues, in an Everettian context. Perhaps. But if so, a case needs to be made out. The significance of the example is that it suggests at least two ways in which such a case might be weak -- two ways in which, at least arguably, a reasonable decision rule in an Everett world would not follow a Born-weighted MEU model.

\section{Branching and distributive justice}

Decision in the Everett world concerns the welfare of a \emph{group} of future individuals --  one's future descendants, or more generally one's `objects of concern', in all future branches.  This fact is highlighted by Greaves'  approach, but clearly true for the subjective uncertainty approach as well. (As we saw, subjective uncertainty might make it \emph{possible} to be selfish, but can't make it rationally \emph{obligatory}. And my disquiet in the Bondi example isn't offset by certainty that it's me who gets the legs.)
The challenge is that 
rational decision in such a context -- a context involving group welfare -- seems in some respects fundamentally different from any weighted sum model.

As I noted, this is a familiar point in discussions of distributive justice.
The problem with  a weighted-sum allocation of goods to a group is that it always permits a large cost to one individual to be offset by small gains to others.  A principle of maximising such a sum thus conflict with plausible principles of justice: 
`Pleasure for some should not knowingly be gained at the cost of pain for
another.'

The plausibility of some such principle is highlighted by the famous Trolley Problem, in which we are asked whether it would be acceptable to kill one person -- e.g., in one well-known version, a fat man sitting on a bridge, who might be pushed into the path of a runaway tram, thus killing him but saving the lives of several passengers -- in order to benefit others. To a remarkable degree, ordinary humans seem to agree that this would not be morally acceptable.\footnote{Another version, even closer to Legless at Bondi, is Transplant:
\begin{quote}
Imagine that each of five patients in a hospital will die without an organ transplant. The patient in Room 1 needs a heart, the patient in Room 2 needs a liver, the patient in Room 3 needs a kidney, and so on. The person in Room 6 is in the hospital for routine tests. Luckily (for them, not for him!), his tissue is compatible with the other five patients, and a specialist is available to transplant his organs into the other five. This operation would save their lives, while killing the ``donor''. There is no other way to save any of the other five patients. \ldots\ [W]ith the right details filled in, it looks as if cutting up the ``donor'' will maximize utility, since five lives have more utility than one life. If so, then classical utilitarianism implies that it would not be morally wrong for the doctor to perform the transplant and even that it would be morally wrong for the doctor not to perform the transplant. Most people find this result abominable. They take this example to show how bad it can be when utilitarians overlook individual rights, such as the unwilling donor's right to life. (Sinnott-Armstrong 2006)
\end{quote}}

If we approach the choices facing Everettian agents in this frame of mind, viewing them as problems of group welfare, then some 
attractive decision principles might  include the following:

 \begin{enumerate}
\item First, do no harm: we should try to establish a baseline, below which we don't knowingly allow our descendants to fall (at least not simply for the sake of modest advantage to others). \pause
\item Second, unequal `good luck' seems a lot less objectionable than unequal `bad luck'. An Everettian might be  inclined to play lotteries, for example, accepting that most of his descendants will make rather trivial sacrifices, in order to make it certain that one will hit the jackpot. 
\item Third, we might trade off `quantity' for `quality', as in quantum Russian roulette (Squires 1986, 72).
\end{enumerate}
This isn't intended as a definitive list  -- it is far from clear there could be such a thing. It is simply intended to illustrate that agents who approach Everettian decision problems in the spirit of Greaves' caring measure, with the welfare of descendants uppermost in their minds, need not see matters entirely, or primarily, in terms of an MEU model. There are other intuitive appealing decision principles available in such cases, of a much more qualitative nature. An argument for the unique reasonableness of the \D ian global preference would need to convince us that these contrary intuitions are guilty of some deep error.

\subsection{Objections to the distributive justice analogy}

\subsubsection{What about the axioms of rationality?}

One objection appeals to the plausibility of the decision-theoretic axioms. Granted, there are other possible preferences for the global state, other than a preference for maximising the Born-weighted sum of in-branch utilities. (Granted, too, perhaps, that this is an important difference, in principle, from the classical case.) But doesn't the intuitive appeal of the axioms still provide a sense in which such a preference would be \emph{unreasonable?}

There are two ways to counter this objection, a direct way and an indirect way. The direct way is to give examples of ways in which intuitively reasonable global preferences can conflict directly with some of the axioms in question. Again, the analogy with classical problems of distributive justice is likely to prove helpful. Suppose I have a choice of leaving $\$1000$ to each of my seven children, or $\$1100$ to six of them and $\$1000$ to the last. The latter choice \emph{dominates} the former, in that no child is worse off and some are better off. But is it a better choice? Many of us would say that it is worse, because it is unjust.\footnote{For some different axiom-busting considerations, see Lewis 2005, \S 4.}

The indirect way  to counter the objection is to point out that  the appeal to the axioms makes assumptions about an agent's utility function that simply beg the question against the possibility of agents of the kind relevant here: agents who have preferences (of a non-MEU form) about the global state. As I've explained, the Oxford approach amounts to trying to mandate a particular global preference as a principle of rationality. What's wrong with an attempt to appeal to the axioms to rule irrational an alternative global preference is that the axioms make assumptions about utilities, about what an agent cares about, that are simply inapplicable to the kind of agent my objection has in mind -- an agent for whom the payoff lies in an Everettian version of distributive justice, for example.

\subsubsection{A difference between Everett and distributive justice?}
\begin{figure}[htbp]
\begin{center}
\includegraphics[width=4.8in]{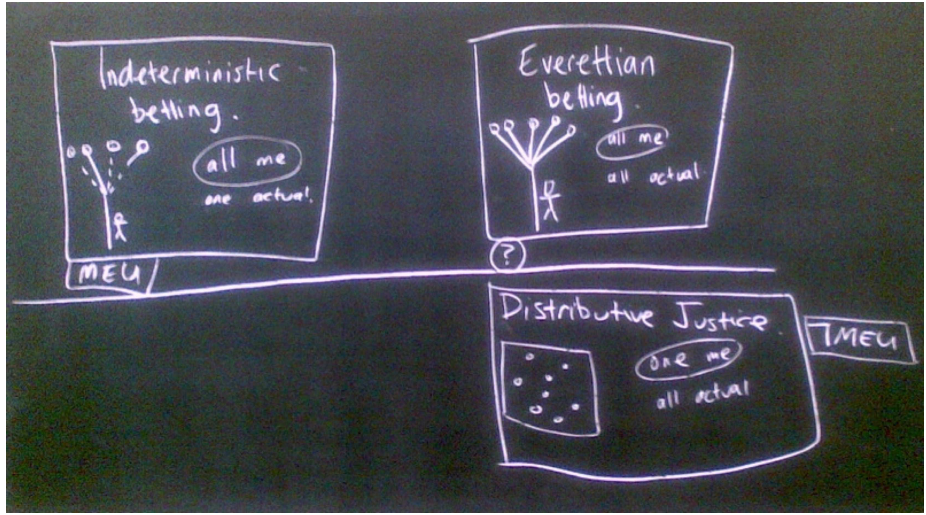}
\caption{Greaves on Everett and Distributive Justice}
\label{default}
\end{center}
\end{figure}\vspace{6pt}

\noindent In a panel discussion at the Many Worlds@50 meeting, Greaves (2007c) suggested a way of drawing a distinction between Everettian decisions and those faced in classical problems of distributive justice. Her idea (see Fig.~1) is that there are two dimensions on which we may compare and contrast the Everettian 
situation to one-branch decision under uncertainty, on one side, and cases of distributive justice on the other. One dimension -- the one emphasised in my objections -- tracks the number of individuals involved. In this dimension, we have one recipient in cases of classical uncertainty, and many recipients both in Everett and in cases of distributive justice. The other dimension -- the one that Greaves recommends that we should emphasise instead -- tracks the issue as to whether all the individuals involved are `me' (i.e., the agent). Greaves suggests that it is the fact that all the beneficiaries are `me' that unifies classical uncertainty and Everett; and the fact that not all are `me' that makes distributive justice different (so that MEU doesn't apply, in those cases). 

As I've set things up above,  however, we've already taken the `me' out of the picture, for other reasons. (The basic reason was that only `me' relevant to classical subjective decision theory is the `me at the time of choice'.) So it simply isn't available to do the work that Greaves wants it to do, in the present context.\footnote{Could we replace it, for Greaves' purposes,  with my notion of `locus of concern'? No, for it simply isn't true, of classical distributive justice problems, that there is only one person in the group whose welfare is `really' my concern. (Recall the case of arranging one's affairs to benefit a group of children.)} 

\subsubsection{What about \emph{weight?}}

Finally, it might be objected that the distributive justice challenge simply ignores the branch weights  -- the crucial disanalogy between the Everett case and classical problems of distributive justice is precisely that in the former case, it is not true that all `concernees' are created equal, at least from the decision-maker's perspective. After all, that's what weight \emph{is,} on Greaves' view: a measure of rational degree of \emph{care.} Don't we simply beg the question against the view, if we insist on the analogy with classical distributive justice?

This brings us to a crucial issue. What \emph{kind} of consideration could it be, in the Everett picture, that would make it rational to give more consideration to some of our loci of concern than others, in the way proposed. I'll approach this issue by considering the analogous question in the classical context.

\section{Why should weight matter?}

\subsection{The Credence-Existence Link}

Consider the initial version of Legless at Bondi (only my leg is cloned). Suppose I survive the operation, as planned, with two good legs. Why don't I care about the unfortunate `possible me', whom the operation left with no legs at all? Because, happily, he doesn't exist. (I have no duty of care to a Man Who Never Was.) 

Why didn't I care about him (much) before the operation, when I wasn't certain that he didn't exist? Because I was \emph{very confident,} even then, that he wouldn't exist. And in the limit, being very confident equates simply to believing. At least to a first approximation, then, we can say that I ignored him because I believed -- rightly, as it turned out -- that he didn't exist.\footnote{And note that if my belief had proved false, I might be culpable for that, but not for what I did in the light of it -- given what I believed, I was right to ignore him.}
In the classical case, then, the justification for giving preference to higher weighted alternatives goes something like this. We give (absolute) preference to actual things over `merely possible' versions of the same things, and the weights simply reflect our degrees of confidence about which things are actual. (Call this the \emph{Credence-Existence Link.})\footnote{True,  there are some subtleties about the notion of belief as a limit of credence. One way to unpack the idea a little further, sticking closely to the pragmatic understanding of credence, goes something like this.  In practice, any assignment of degrees of belief has a resolution limit -- an `epistemic
grain', beneath which differences are of no practical significance. At this limit, low probability equates to zero, for the purposes at hand.
I act \emph{just as if} I believe that the low-weight alternatives don't exist.} 

This doesn't work in the Everett case: if 
well-weighted pleasure comes at the expense of low-weighted pain,
then it is \emph{real} pain, despite its weight. What Greaves needs is a measure which has the same \emph{normative} implications as `confidence in  non-existence', while being entirely compatible with `certainty of existence'.\footnote{Greaves proposes that existence might come by degrees, though reality doesn't: 
\begin{quote}
Lockwood's \ldots\
talk of a `superpositional dimension', and/or Vaidman's \ldots\ suggestion 
that we speak of the amplitude-squared measure as a `measure of existence', are 
somewhat appropriate (although we are not to regard lower-weight successors 
as \emph{less real,} for \emph{being real} is an all-or-nothing affair -- we should say instead 
that there is \emph{less} of them).' (Greaves 2004, 30)
\end{quote} 
For my part, I can't see how such a distinction could make a difference here. Why should my concern for another's welfare depend on whether he \emph{exists} (and hence on the degree to which he exists, on this view), but not simply on whether he is \emph{real?} And in any case, the kind of `Do no deliberate harm' principles which weight needs rationally to trump, to meet the objections above, are insensitive to quantity in the sense of frequency. What could make them sensitive to quantity in the sense of degree of existence?} That's what an analogue of classical uncertainty would look like.

\subsection{What alternative is there?}

Greaves suggests that there is no realistic alternative to a caring measure based on the Born weights. But we need to distinguish two questions.
First, is there any rival MEU approach -- i.e., any rival to the Born measure, if Everettian decision theory is agreed to an analogue of classical decision under uncertainty?
And second, is there any rival to the MEU approach itself?
We need to answer the second question, before we can take refuge in a negative answer to the first. 

We've seen that there do seem to be non-MEU alternatives, modelled in part on problems of distributive justice. It is no answer to this challenge, obviously, to claim that there are no alternatives \emph{within} the MEU framework. Moreover, Greaves herself notes another possible positive answer to the second question, viz., the nihilist option, that rational action is simply incoherent in the Everett world:
\begin{quote}
[I]t could turn out that there were no coherent strategy for 
rational action in an Everettian context. \ldots\  As Deutsch 
notes, this possibility cannot be ruled out of court: 
\begin{smallquotation}
\noindent It is not self-evident \ldots\   that rationality is possible at all, in the presence of 
quantum-mechanical processes---or, for that matter, in the presence of electromagnetic or any other processes. (Deutsch, 1999, p.~3130) (Greaves 2004, 432)
\end{smallquotation}
\end{quote}
The `what alternative is there?'~plea can hardly have much force, presumably, while nihilism waits in the wings (drawing strength from the difficulties that the DW approach has in explaining why weight should have normative significance).\footnote{In the light of \S 5.1, nihilism can only be view that there is no \emph{preferred} rational strategy. As we saw, suitable preferences about the global state can certainly determine a rational strategy (without uncertainty). So nihilism needs to be understood as the view that there is no rational constraint on global preferences themselves.}

Greaves notes that the most commonly suggested rival to the Born measure is what she calls \emph{Egalitarianism}, or \emph{Naive Counting:} a view that tries to treat all branches equally. Following Wallace, however,  Greaves (2007b, 120) claims that this view turns out to be incoherent,  in a decoherence-based version of the Everett picture, because the number of branches is not well defined: 

\begin{quote}
Why does naive counting break down in the decoherence approach? The core of the 
problem is that naive counting \ldots\  presupposes the existence of a piece of structure that 
is not in fact present in the theory. 
\end{quote}
However, I think that Greaves and others have failed to distinguish two versions of Egalitarianism, only one of which is aptly called Naive Counting (and subject to this objection).\footnote{Cf. Lewis's (2005, 14--15) distinction between the Average Rule and the Sum Rule.} The version Greaves has in mind is a view that accepts that rational decision in the Everett world should properly be an analogue of classical decision under uncertainty, but simply proposes a rival measure -- a measure that in the finite case in which there are $N$ branches, gives each branch a weight of $1/N$. Grant that this turns out to be incoherent, for the reason Greaves describes.

The rival form of Egalitarianism (`Outcome Egalitarianism') simply rejects the attempt to make numerical comparisons, treating all non-null \emph{outcomes} as having the same kind of claim to be taken into account. Thus it rejects the idea that there is any sort of comparative weight to be associated with outcomes, with normative significance, whether counting-based or not. Is this view incoherent? If so, it doesn't seem to be because the number of \emph{branches} is ill-defined in a decoherence-based version of the Everett interpretation -- this view embraces that lesson.

\section{Conclusion}

Some brief conclusions. The `Why does weight matter?' challenge has not been met. It is far from clear that there aren't compelling alternatives, or at least supplements, to a \D ian decision policy, for an agent who believes that she lives in an Everett world. (Some of these alternatives are motivated by considerations analogous to those of distributive justice, but perhaps not all -- if quantum Russian roulette appeals, it is for other reasons.)  It is true that these considerations fall a long way short of a unified Everettian decision policy, and in particular, do little to stave off the larger threat of nihilism. But without a explanation as to why weight should have normative significance,  \D ian decision theory does no better in this respect.
Most importantly, there seems  little prospect that a \D ian decision rule can be a constraint of rationality, in a manner analogous to the classical case. The fundamental problem rests squarely on the distinctive ontology of the Everett view: on the fact that it reifies what the one-branch model treats as mere\emph{ possibilia,} and hence moves its own decision rule into the realm of `mere preference'. 
I conclude that the problem of uncertainty has not been solved.\footnote{It is worth noting that this whole issue takes for granted that free choice makes sense in the
Everett world, as much as it does in the one-world case. But there
are reasons to doubt that, too.
  Whenever I form an intention to $\phi$, there's a small `probability' that my intention will be thwarted.
Hence in an Everett world, I know that my `decision' to   $\phi$ produces some descendants who  $\phi$ and some descendants who don't  $\phi$.
 What sense, then -- at least \emph{in advance} of a justification for treating weights as probabilities -- can we make of the notion of an effective choice between games?  (The problem isn't that my choice is \emph{determined,} but that I can't make an effective choice \emph{between alternatives}.) The Everett view thus seems vulnerable to a new kind of argument for fatalism.}

\section*{Acknowledgements}

My interest in these issues dates from a characteristically lucid talk by David Wallace in Konstanz in 2005, which introduced me to the basics of the DW approach. It was further encouraged by the opportunity to discuss them with Hilary Greaves in Sydney in 2006. I'm grateful to the organisers of the Many Worlds@50 Conference at the Perimeter Institute for inviting me to take part in such a fascinating meeting, and to Wayne Myrvold, David Papineau and other participants there, for further discussions. I'm also very much indebted to Guido Bacciagaluppi, Jenann Ismael and especially Peter Lewis,  for helpful conversations in Sydney about previous versions of this material. And I'm grateful to the Australian Research Council and the University of Sydney, for research support.

\section*{Bibliography}

\vspace{6pt}\noindent  Deutsch, D. (1999).
 {Q}uantum {T}heory of {P}robability and {D}ecisions.
 {\em Proceedings of the Royal Society of London\/}~A455,
  3129--3137.

\vspace{6pt}\noindent Greaves, H. (2004).
 Understanding {D}eutsch's Probability in a Deterministic Multiverse.
 {\em Studies in the History and Philosophy of Modern Physics}~35, 423--456.

\vspace{6pt}\noindent Greaves, H. (2007a).
 On the Everettian Epistemic Problem. \emph{Studies in History and Philosophy of Modern Physics}~38, 120--152.

\vspace{6pt}\noindent Greaves, H. (2007b).
Probability in the Everett Interpretation. 
\emph{Philosophy Compass}~2(1), 109--128.

\vspace{6pt}\noindent Greaves, H. (2007c).
Comments in Panel Discusion, \emph{Many Worlds@50 Conference,} Perimeter Institute, 24 September 2007.
[Video and audio available here: \href{http://pirsa.org/07090076/}{http://pirsa.org/07090076/}]

\vspace{6pt}\noindent Lewis, D. (2004). How Many Lives has Schrodinger's Cat? \emph{Australasian Journal of Philosophy}~82,  3--22. 
  
\vspace{6pt}\noindent   Lewis, P. (2005).
 Probability in Everettian Quantum Mechanics.
 Available online from \href{http://philsci-archive.pitt.edu/archive/00002716/}{http://philsci-archive.pitt.edu/archive/00002716/}.

\vspace{6pt}\noindent   Papineau, D. (1996). {M}any Minds Are No Worse Than One.
{\em British Journal for the Philosophy of Science}~47,
  233--241.
  
\vspace{6pt}\noindent   Price, H. (2007). Decisions, Decisions, Decisions: Thoughts About Actions in an Everett World. \emph{Many Worlds@50 Conference,} Perimeter Institute, 23 September 2007.
[Video and audio available here: \href{http://pirsa.org/07090075/}{http://pirsa.org/07090075/}] 
  
 \vspace{6pt}\noindent  Saunders, S. (1998). Time, Quantum Mechanics, and Probability. \emph{Synthese}~114, 
 373--404. 
 
 \vspace{6pt}\noindent  Saunders, S. (2005). What is Probability? \emph{Quo Vadis Quantum Mechanics.} Eds. A.~Elitzur, S.~Dolev \& 
N.~Kolenda. Springer-Verlag.

 \vspace{6pt}\noindent  Sinnott-Armstrong, W. (2006). Consequentialism. \emph{The Stanford Encyclopedia of Philosophy (Winter 2006 Edition),} Edward N. Zalta (ed.). Accessible online at  \href{http://plato.stanford.edu/archives/win2006/entries/consequentialism/}{http://plato.stanford.edu/archives/win2006/entries/consequentialism/}.

 \vspace{6pt}\noindent  Squires, E. (1986). \emph{The Mystery of the Quantum World.} Bristol: Adam Hilger.
 
  \vspace{6pt}\noindent  Wallace, D. (2002). Quantum Probability and Decision Theory, Revisited. Available online at \href{http://www.arxiv.org/abs/quant-ph/0211104}{quant-ph/0211104}.
  
 \vspace{6pt}\noindent  Wallace, D. (2003).
{E}verettian rationality: defending {D}eutsch's Approach to
  Probability in the {E}verett interpretation.
{\em Studies in the History and Philosophy of Modern Physics}~34, 415--439.

\vspace{6pt}\noindent Wallace, D. (2006). Epistemology Quantized: Circumstances in Which We Should Come to  Believe in the Everett Interpretation.  \emph{British Journal for the Philosophy of Science}~57, 655--689.
  
\vspace{6pt}\noindent Wallace, D. (2007).   Quantum Probability from Subjective Likelihood: Improving on Deutsch's Proof of the Probability Rule.  \emph{Studies in History and Philosophy of Modern Physics} 38, 311--332.

\end{document}